\documentclass[nonacm]{acmart}
\usepackage{graphicx}
\usepackage{subcaption}
\usepackage{url}
\usepackage{hyperref}

\AtBeginDocument{%
  \providecommand\BibTeX{{%
    \normalfont B\kern-0.5em{\scshape i\kern-0.25em b}\kern-0.8em\TeX}}}

\long\def\omitit#1{}

\begin{document}

\title{AI-Augmented Brainwriting: Investigating the use of LLMs in group ideation}

\author{Orit Shaer}
\authornotemark[1]
\email{oshaer@wellesley.edu}
\orcid{0000-0002-0515-2957}

\affiliation{%
  \institution{Wellesley College}
  \streetaddress{106 Central st.}
  \city{Wellesley}
  \state{MA}
  \country{USA}
  \postcode{02481}
}

\author{Angelora Cooper}
\email{acooper5@wellesley.edu}
\orcid{0000-0003-0758-7740}
\affiliation{%
  \institution{Wellesley College}
  \streetaddress{106 Central st.}
  \city{Wellesley}
  \state{MA}
  \country{USA}
  \postcode{02481}
}

\author{Osnat Mokryn}
\orcid{0000-0002-1241-9015}
\email{omokryn@is.haifa.ac.il}
\affiliation{%
  \institution{University of Haifa}
  \streetaddress{199 Abba Khushi Ave.}
  \city{Haifa}
  \country{Israel}}

\author{Andrew L. Kun}
\email{andrew.kun@unh.edu}
\orcid{0000-0001-9756-7748}
\affiliation{%
  \institution{University of New Hampshire}
  \city{Durham}
  \state{NH}
  \country{USA}
}

\author{Hagit Ben Shoshan}
\orcid{0009-0007-5945-695X}
\email{hagits@gmail.com}
\affiliation{%
  \institution{University of Haifa}
  \streetaddress{199 Abba Khushi Ave.}
  \city{Haifa}
  \country{Israel}}

\renewcommand{\shortauthors}{Shaer and Cooper, et al.}

\begin{abstract}
The growing availability of generative AI technologies such as large language models (LLMs) has significant implications for creative work. This paper explores twofold aspects of integrating LLMs into the creative process – the divergence stage of idea generation, and the convergence stage of evaluation and selection of ideas. We devised a collaborative group-AI Brainwriting ideation framework,  which incorporated  an LLM as an enhancement into the group ideation process, and evaluated the idea generation process and the resulted solution space. To assess the potential of using LLMs in the idea evaluation process, we design an evaluation engine and compared it to idea ratings assigned by three expert and six novice evaluators. Our findings suggest that integrating LLM in Brainwriting could enhance both the ideation process and its outcome. We also provide evidence that LLMs can support idea evaluation.  We conclude by discussing implications for HCI education and practice.

\end{abstract}

\ccsdesc[500]{Human-centered computing~User studies}
\ccsdesc[500]{Human-centered computing~Collaborative interaction}

\keywords{LLM, Brainwriting, human-AI collaboration}



\maketitle

\section{Introduction}

The increasing availability of generative AI technologies \cite{nytimesWhatsFuture} such as large language models (LLMs) \cite{openaiGPT4,LLaMAOpenMetaResearch,GoogleBardChat2023} and image generators \cite{Midjourney2022,DALLE22022,StableDiffusionOnline2022} 
has significant implications for creative work~\cite{hbr2022GenerativeAI,Olsson21}. Given their wide adoption \cite{forbesCouncilPost}, it is critical to investigate the merits and limitations of integrating such tools into the creative process through new forms of co-creation.

Recent work has begun to explore how co-creation with generative AI could be used for interaction design~\cite{Schmidt23} and what co-creation practices might look like for problem solving~\cite{salikutluk2023interacting}, ideation \cite{Verheijden23,Tholander23,Kim23}, prototyping, making, and programming~\cite{Jonsson22,Reddy22,Wang_Popblend,vaithilingam2022expectation}. Emerging theories about posthumanism, post-human, and more-than-human interaction design \cite{Giaccardi20, wakkary2020nomadic,Wakkary2021,Homewood21} provide further context for human-AI co-creation activities by highlighting possibilities to distribute agency in design among humans and non-humans.

The overarching research question we are interested in is 
how LLMs can contribute to enhancing the human creative thought process through new forms of co-creation for groups. 
In this paper, we take a step toward exploring this question by focusing on the use of LLMs in a specific type of a creative ideation process for groups: \textit{Brainwriting}~\cite{Wilson2013}. Brainwriting derives from brainstorming \cite{osborn1953applied}, which is a structured technique for group ideation. During a successful group brainstorming session, participants draw on each other’s ideas and pre-existing knowledge to combine ideas in new ways~\cite{hargadon2003breakthroughs}. Despite the perception that groups are more productive at brainstorming, a greater number of ideas and better quality ideas are often found in individual brainstorming~\cite{diehl1987productivity}. This is because individuals working alone tend to consider many different potential solutions, while group members working together often consider fewer alternative solutions due to peer judgment, free riding, and production blocking~\cite{hymes1992unblocking}.   

Brainwriting~\cite{Wilson2013} is an alternative or a complement to face-to-face group brainstorming, which aims to address these shortcomings. It begins with asking participants to write down their ideas in response to a prompt before sharing their ideas with others. After writing ideas in a parallel process, participants review others’ ideas and add new ones. The number of ideas generated from Brainwriting often exceeds face-to-face brainstorming because of the more inclusive parallel process \cite{Wilson2013}.  
With the capability of LLMs to generate new content, several commercial products have integrated LLMs support for Brainwriting in their products (e.g.~\cite{miroAI2023, nodaNodaMind}).

This paper explores twofold aspects of integrating LLMs into a group Brainwriting ideation process – the divergence stage of idea generation, and the convergence stage of evaluation and selection of ideas. Specifically, our investigation focuses on the following research questions:

\begin{description}
\item [RQ1:] Does the use of an LLM during the divergence stage of collaborative group Brainwriting enhance the idea generation process and its outcome?











\item [RQ2:] How can LLMs assist to evaluate ideas during the convergence stage of a collaborative group Brainwriting process? 

\end{description}

To explore these questions we devised a collaborative group-AI Brainwriting ideation framework, which incorporated an LLM as an enhancement into the group ideation process. We evaluated the use of the framework during the divergence stage for idea generation and the resulting solution space (RQ1) by integrating it into an advanced undergraduate course on tangible interaction design. The course seeks to expose students to novel human-AI co-creation processes within tangible interaction design~\cite{Shaer2023}, and to prepare them to engage with emerging LLM-based interaction design methods~\cite{Schmidt23}. We conducted the evaluation with 16 students using both qualitative and quantitative methods. 

To assess the potential of using LLMs in the divergence stage of group Brainwriting for idea evaluation (RQ2), we designed an LLM evaluation engine, which rates ideas based on three criteria: \textit{Relevance} -- the extent to which the idea is connected to the problem statement, \textit{Innovation} -- how original and creative the idea is, and \textit{Insightfulness} -- the extent to which the idea reflects a profound and nuanced understanding of the problem statement. We then compared the ratings produced by the LLM evaluation engine to ratings assigned by three expert and six novice evaluators. 

This paper contributes to the HCI field by expanding the pedagogical frameworks and offering new AI-augmented tools for educators and novice designers, as well as by providing empirical insights into the challenges and opportunities of incorporating AI into collaborative ideation. Specific contributions include:  1) a collaborative group-AI Brainwriting ideation framework which enhances both divergent and convergent stages; 2) an LLM idea evaluation engine, which rates idea quality based on relevance, innovation, and insightfulness;  3) empirical insights into how the Brainwriting participants who are novice designers engage with and perceive the process of group-AI Brainwriting; 4) evidence that integrating the use of LLM into Brainwriting could enhance both the ideation process and its outcome; 5) evidence that LLMs can assist users in idea evaluation; 6) finally, we discuss merits and limitations of integrating LLMs into a collaborative brainwriting ideation process for both HCI education and practice.

In the following we describe the designed framework, our methods and findings. We begin with related work.
 
\section{Related Work}

\subsection{Structured Approaches to Ideation}
Structured approaches to generating, refining, and evaluating ideas play a crucial role in creative processes across domains.  Collaborative ideation approaches include techniques such as brainstorming \cite{osborn1953applied}, Brainwriting \cite{Wilson2013}, and Six Thinking Hats \cite{Bono1999}. Research indicates that collaborative approaches for ideation could lead to more creative solutions because when people are exposed to different perspectives, they might be inspired to explore new connections through diverse ideas~\cite{Herringinspired, Dowprototyping,Leeexamples,Siangliulue}.  

To leverage diversity of ideas, several online platforms for large-scale ideation allow users to share their ideas and to explore ideas shared by others. However, in order to expose users to those ideas that are creative and potentially inspiring, such systems need to implement methods to select and present creative and diverse ideas \cite{Siangliulue}. HCI and CSCW research have demonstrated various crowd-based and algorithmic approaches for addressing this challenge~\cite{Siangliulue}.

In this paper, rather than focusing on large-scale ideation, we explore ways to enhance small groups (3-4 people) ideation through the use of LLMs. Brainstorming~\cite{osborn1953applied} is one of the most widely adopted techniques for generating creative ideas within groups~\cite{Devine1999}. However, there are several known barriers which limit the effectiveness of group brainstorming in producing a high number of high quality creative ideas~\cite{STROEBE2010157}, including peer judgment, group thinking, free riding, and production blocking - when group members wait for their turn before sharing an idea~\cite{diehl1987productivity}. It is also shown that group members tend to overestimate their group productivity and creativity~\cite{paulus1993social}. 

Brainwriting \cite{Wilson2013}, is an alternative or complementary method to face-to-face group brainstorming, which aims to address these shortcomings through a parallel rather than sequential process. While there are several variations of the process \cite{HeslinBrainwriting}, generally, in a Brainwriting session, participants are asked to write down their ideas in response to a prompt before sharing their ideas with others. After writing ideas in a parallel process, after participants work silently on writing their ideas, participants review others’ ideas and then add new ones by either individually writing additional ideas or through discussion and collaboration. The number of quality ideas generated from Brainwriting sessions often exceeds face-to-face brainstorming because the process mitigates the barriers posed from brainstorming through a more inclusive parallel process \cite{paulus2000idea}, however it is important to consider context and adjust the process for the specific group characteristics \cite{HeslinBrainwriting}.
In recent years, online visual workspaces such as Miro~\cite{miroProductOverview2023}, ConceptBoard~\cite{SecCollaborationConceptboard2023} 
and Mural~\cite{MuralVisualWorkPlatform2023} offer support and template for remote and co-located Brainwriting processes. With the increasing capability of LLMs to generate new content, such services have integrated LLMs functionality as part of their products. However, there is little knowledge about the merits and limitations of integrating LLMs into ideation processes. Shin et al. led a CHI 2023 workshop to explore the integration of AI in human-human collaborative ideation \cite{Shinworkshop}. Our goal is to add to the emerging body of knowledge on collaborative group-AI ideation.


\subsection{Human-AI Co-Creation}
Co-creation, where humans and machines work together to create new artifacts or solve a problem, is not new. The origin of computer-aided design (CAD) could be traced back to the pioneering Sketchpad system~\cite{UCAM-CL-TR-574}, which was created by Ivan Sutherland as part of his 1963 doctoral dissertation. The system, among other breakthrough innovations in computer graphics, human-computer interaction, and object-oriented programming, demonstrated that a user and a computer could “converse rapidly through the medium of line drawings”~\cite{10.1145/1461551.1461591}.  Modern CAD practices, which include generative design, have been used by designers to explore and expand their design space~\cite{10.1145/3196709.3196813, McCormack_Dorin_2014}. 


With the emerging availability of generative AI models and tools, recent work has begun to explore how co-creation with AI models, which are not domain-specific, could be used for interaction design and what co-creation practices with generative AI tools might look like for ideation~\cite{hbr2022GenerativeAI,Tholander23, Verheijden23, Kim23}, persona creation~\cite{goel2023preparing} prototyping, making, and programming~\cite{10.1145/3373644, Jonsson22,Reddy22}. 

Most relevant to this case study is a small scale study conducted by Tholander and Jonsson \cite{Tholander23} with experienced designers, which examines how large language models and generative AI can support creative design and ideation. Their findings highlight both opportunities and challenges in integrating and using GPT-3 and Dall-E by experienced designers. The work we present in this case study, extends previous work by shedding light on how students who are \textit{novice designers}, interact with and perceive the results of ideas co-created with LLMs. 

These examples of co-creation could be contextualized within emerging theories about post-humanism, post-human, and more-than-human interaction design~\cite{Giaccardi20, wakkary2020nomadic,Wakkary2021,Homewood21}. These theories consider alternatives to human-centered design, challenging the assumption of the “human at the center of thought and action”~\cite{Wakkary2021} by arguing that agency is distributed among humans, non-humans, and the environment. In response to these theories, van Dijk cautions that post-human design could obscure the important fact that non-humans agents such as AI technology are trained upon and imports traditional, humanist forms of logic and language, which in turn might taint post-human design with their humanist roots and biases~\cite{vanDijk2020}.

\subsection{Approaches for Evaluating Ideas}
Dean and colleagues provide a framework for evaluating ideas \cite{Dean2006IdentifyingQN}.  The framework has four dimensions — \emph{novelty}, \emph{workability} (also called \emph{feasibility}), \emph{relevance}, and \emph{specificity}. The framework allows a systematic evaluation of the \textit{quality} of ideas across studies, using common definitions. 

In addition to evaluating the quality of individual ideas, there are also important reasons to evaluate the \textit{quantity} of ideas, which an ideation process generates. This is because people are more likely to find good ideas when choosing from many ideas rather than when only a few are available - in the case of ideation, more is better \cite{johansson2004medici}. For example, there is evidence that having access to more AI-generated ideas improves story-writing \cite{doshi2023generative}. The selection of winning ideas - those ideas that really make a difference - means that when ideas generated by an individual or a team are evaluated, the \textit{average} quality of these ideas is less interesting - after all, as Girotra et al. argue, having a few (or even one) great idea is much better than having many average ideas \cite{girotra2023ideas}. Setting such a high importance on high quality ideas is especially reasonable for cases where there is a single ideation event. 

While the above approaches to idea evaluation are most often associated with humans evaluating ideas, there is also an opportunity to use AI to evaluate ideas. This approach holds the promise of increased speed of idea evaluation, as well as the opportunity to develop human-AI collaborative teams where the AI could support the creative efforts of humans by providing feedback. Thus, researchers have already explored the use of AI to creativity in drawing \cite{Cropley2023FitForPurpose}, and in this work we explore using LLM to evaluate the written ideas generated by teams comprising of humans and another LLM. Domonik shows that AI evaluation could also improve human ideation by reducing evaluation apprehension - the situation where a human will withhold an idea for fear of being evaluated negatively \cite{siemon2023let}. 

\section{Collaborative Group-AI Brainwriting Framework Design}

Our investigation focuses on designing and evaluating a framework for Group-AI Brainwriting. 
The collaborative Group-AI design we were aiming for is one of enhancement, in which  during the \textit{divergence phase}, the group prompts the AI only \textit{after} a first phase of Brainwriting.
Paulus and Yang~\cite{paulus2000idea} suggested a two-phase process for the ideation process, where in the second phase participants recall ideas from the first phase, thus promoting attention and cognitive stimulus. Borrowing from their observation, we design the collaborative group-AI ideation process as a multi-phase process. In the divergence stage, group members first generate their own ideas and add them to a shared online whiteboard. Then, group members review and interact with their collective ideas while prompting an LLM for new ideas that will enhance their initial set of ideas. 

In the convergence stage, group members evaluate the ideas through discussion and narrow the list of ideas to a few selected chosen ideas, which they enhance through the use of an LLM. 
Our investigation seeks to examine the feasibility of expanding the use of LLMs in this stage to assist group members to evaluate their ideas. We devised and evaluated a method for an LLM-based evaluation engine (using GPT-4). 

\begin{figure*}
    \centering
    \includegraphics[width=.98\textwidth]{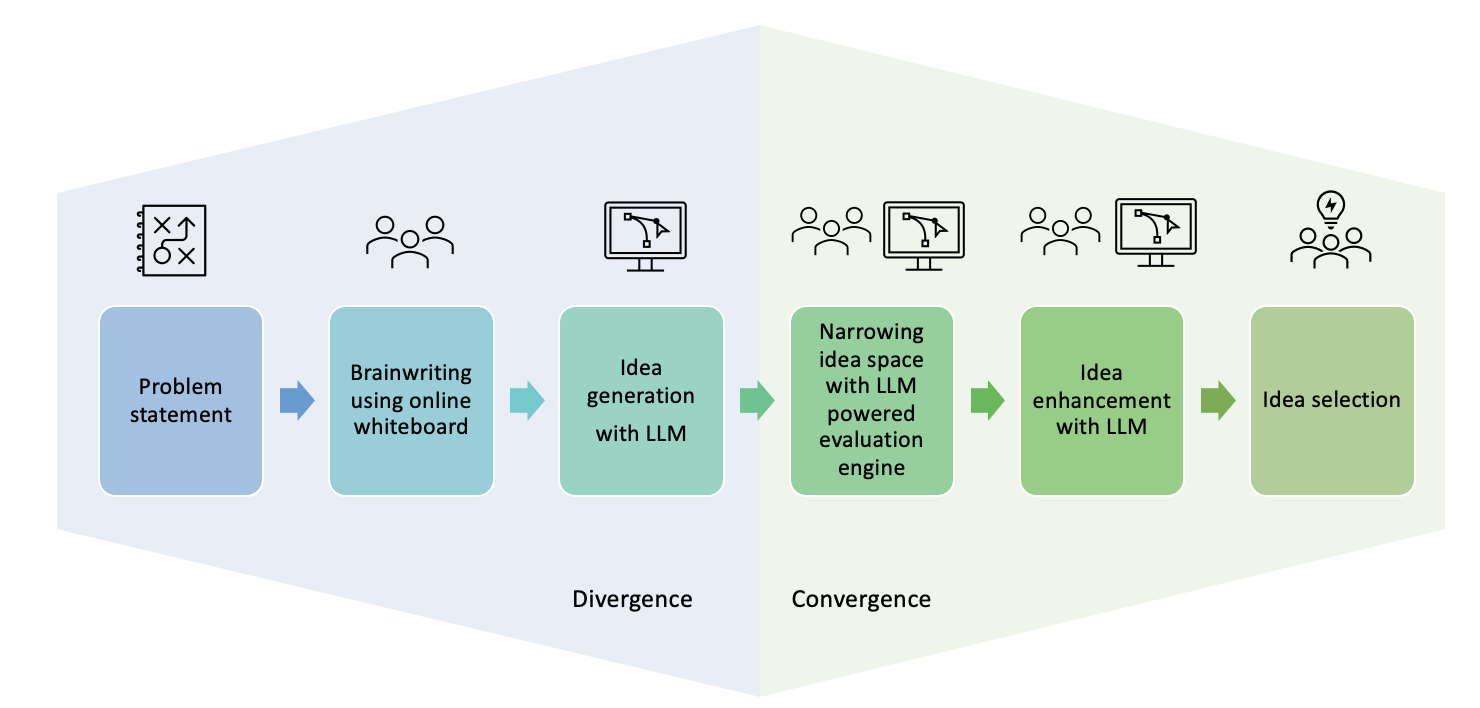}
    \caption{Collaborative Group-AI Brainwriting Process}
    \label{bf}
    \Description{Collaborative Group-AI Brainwriting Process}
\end{figure*}
Figure~\ref{bf} illustrates our proposed collaborative group-AI Brainwriting framework. Following, we describe the elements of this framework. 

\subsection{Brainwriting Divergence stage}
\subsubsection{Phase 1: Brainwriting using Conceptboard}
\begin{figure*}[ht!]
    \centering
    \begin{subfigure}[b]{.95\textwidth}
        \centering    \includegraphics[width=0.9\textwidth]{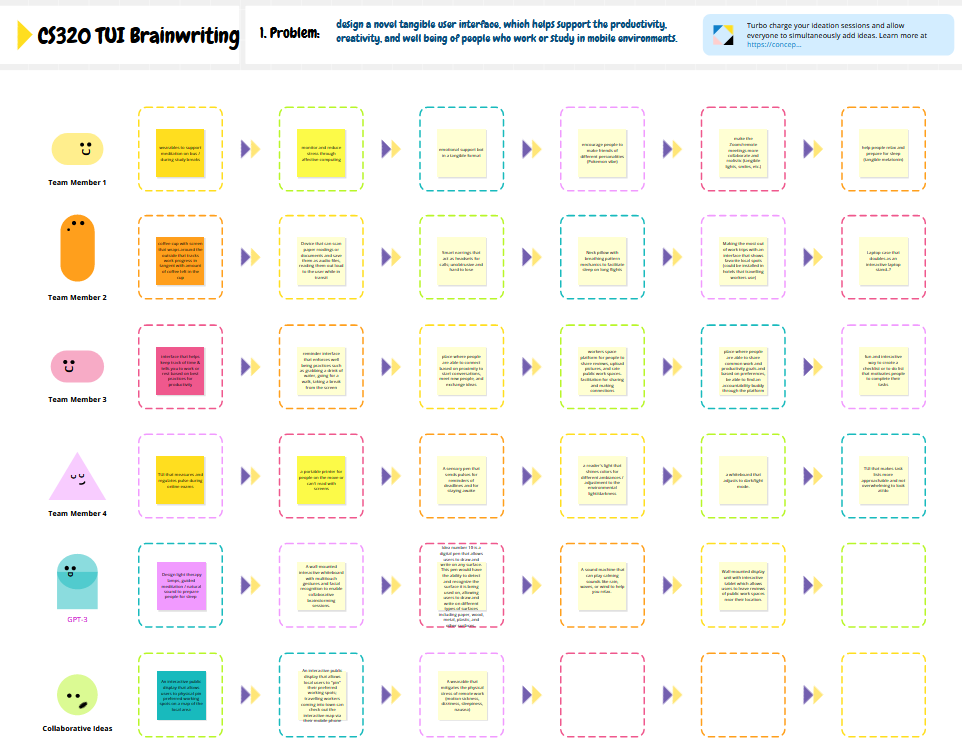}
        \caption{Central area of the Conceptboard, containing the ideas produced during the Brainwriting session}
        \label{fig:top}
    \end{subfigure}
    
    \begin{subfigure}[b]{0.48\textwidth}
        \centering
        \includegraphics[width=.9\textwidth]{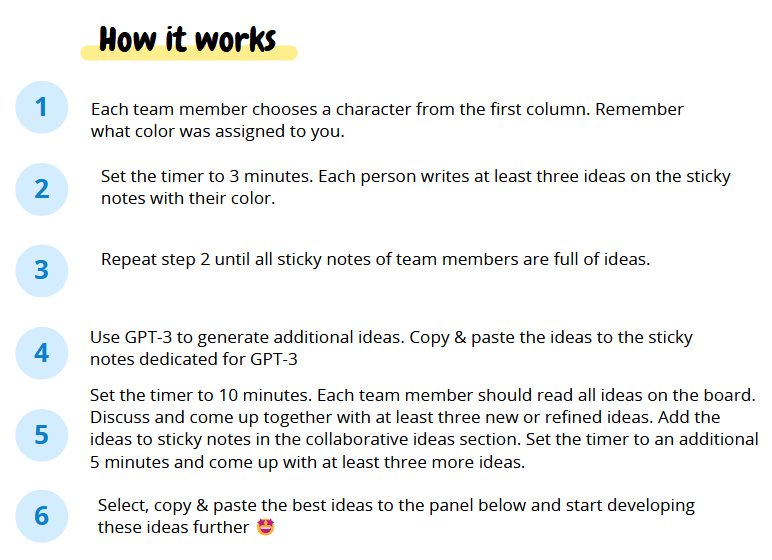}
        \caption{Outline of the process}
        \label{fig:bottom_left}
    \end{subfigure}
    \hfill
    \begin{subfigure}[b]{0.48\textwidth}
        \centering
        \includegraphics[width=0.8\textwidth]{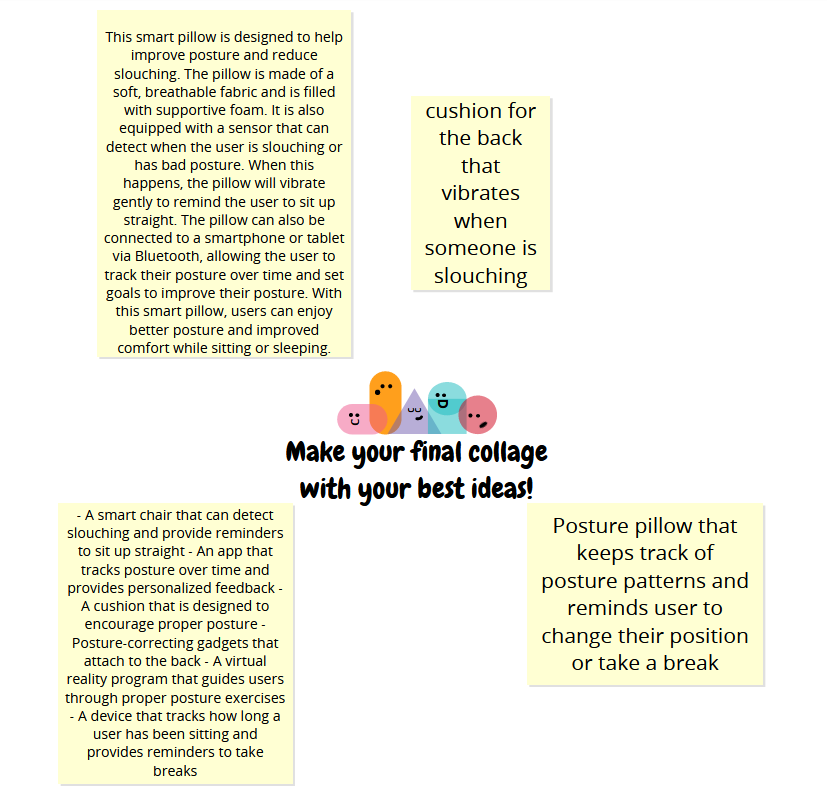}
        \caption{A final set of ideas, chosen from the ideas in the central area}
        \label{fig:bottom_right}
    \end{subfigure}
    
    \caption{The three main areas of Conceptboards used by teams during the Brainwriting session.}
    \label{fig:conceptboard}
    \Description{The three main areas of a Conceptboard used by a team during the Brainwriting session}
\end{figure*}

We modified the Brainwriting process \cite{Wilson2013} so that group members sit together as a team around a shared table, but write their ideas individually, in parallel, on an online whiteboard called Conceptboard \cite{SecCollaborationConceptboard2023}. 
The Conceptboard template we use is based on the Conceptboards remote Brainwriting template \cite{ConceptboardBrainwriting2023}. The problem statement for the Brainwriting session is written at the top of the board. Participants are instructed to each select a color on the board, set a timer for 3 minutes as a group, and use that time so that each group member write at least three ideas relevant to the problem statement and place them on the board using colored coded sticky notes. Then participants are asked to repeat this process until each group member wrote at least six ideas.
Figure~\ref{fig:conceptboard}(a) shows the instructions given to participants. Figure~\ref{fig:conceptboard}(b) shows the modified Conceptboard template we used for the Brainwriting activity, populated with ideas generated by one of the student teams in our study.
Each group worked on a separate Conceptboard. 
 
\subsubsection{Phase 2: Enhancing Ideas with an LLM}
In here,  each group is required to use an LLM (OpenAI Playground GPT-3) to generate additional ideas. Participants are encouraged to iterate on their LLM prompts and are exposed, prior to the Brainwriting session, to overview materials on prompt engineering. The generated ideas are copied and pasted into sticky notes on the board. 
We modified the original Brainwriting template offered by Conceptboard to reflect this new framework for Brainwriting with LLM for the enhancement of ideas.

At this stage the groups were instructed to review all initial ideas, discuss them, and develop together, with the help of GPT-3, new ideas that add to or build upon the existing preliminary ideas. These ideas are added to an area on the board dedicated to collaborative ideas. 

For this stage of the experiment, we selected GPT-3 due to its free availability, which allowed students the opportunity to access and experiment with it in various contexts. 
\subsection{Brainwriting convergence stage}
\subsubsection{Phase 3: Selecting and developing ideas through discussion}
Participants are instructed to select through discussion the best ideas and copy and paste them to a dedicated area on the board. Then they continue to develop these ideas with the help of an LLM.

\subsection{Can LLMs Help with Convergence? Developing and implementing an LLM Powered Evaluation Engine}
Our goal is to examine the feasibility of using LLMs to assist users in the convergence stage by highlighting the most promising ideas from the overall pool and identifying which ideas do not merit further consideration. For this stage we created an LLM evaluation engine. (The LLM-based evaluation was performed after the conclusion of the Brainwriting exercise and was not used to support the Brainwriting process.) 

Our evaluation engine builds on the approach of Dean et al \cite{Dean2006IdentifyingQN} for evaluating the quality of ideas and uses the dimensions of \emph{novelty} (which we call \emph{innovation}) and \emph{relevance} to evaluate ideas. We chose not to use the dimensions of \emph{workability} and \emph{specificity}, because we envision this tool to be used in early stage ideation, in which neither of these dimensions play a large role; both can (and should) be addressed in subsequent stages of ideation. 
We also introduce an additional dimension that we call \emph{insightfulness}, which is based on the work of Dyer et al. on the origin of innovative ventures \cite{Dyeer_innovative}. We define an \emph{insightful} idea as one that reflects a profound and nuanced understanding of the problem statement. 

Several additional aspects need to be considered in the design of an LLM evaluation engine.
First, there should be no ambiguity in the definition and interpretation of the used scales and evaluation criteria. Users would expect such an engine to communicate its evaluations and using shared definitions and agreed-upon scales. Hence, we define the following requirements: 
\begin{description}
\item [Well-known Scale: ] The engine would use a well-known scale, often used by humans. We chose the use of a Likert scale, with a $[1 \ldots 5]$ evaluation range~\cite{allen2007likert}. 
\item [Well-defined Criteria: ] The engine would be prompted to evaluate ideas according to a well defined set of criteria, which is often used by humans to identify quality, innovative, and creative ideas. We chose to use two criteria from Dean et al.'s evaluation framework \cite{Dean2006IdentifyingQN}: \textit{relevance} and \textit{innovation}. In addition, we chose a third criterion, \textit{insightfulness}, based on Dyer et al.'s research on the origin of innovative ventures \cite{Dyeer_innovative}. Each of these criteria required a clear definition. 

\item [Scale x Criteria Definition: ] Each scale value for each criterion should be well defined, and in detail. 

\end{description} 

\begin{figure}
    \centering
    \includegraphics[width=.8\linewidth]{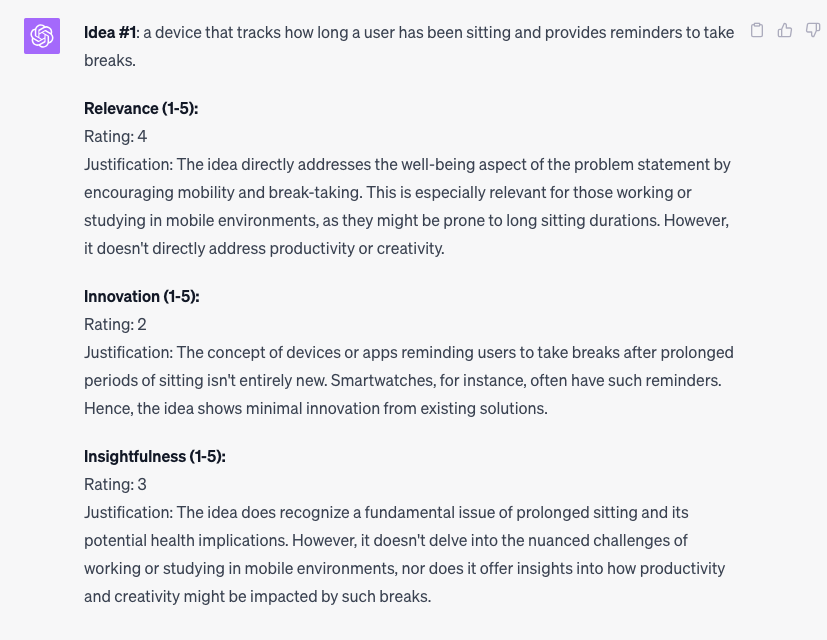}
    \caption{Idea evaluation with GPT-4 using the proposed scales for relevance, innovation, and insightfulness}
    \label{fig:engine}
    \Description{Idea evaluation with GPT-4 using the proposed scales for relevance, innovation, and insightfulness}
\end{figure}

\paragraph{Creating a per scale value and criterion definition. } 
We used the following procedure for developing clear, differentiated, descriptive scale value for each criterion:
\begin{enumerate}
    \item We first developed initial descriptive paragraphs for each criterion - \textit{Relevance, Innovation, Insightfulness}, based on definitions in  existing literature, and created descriptive anchors for each scale value.

    \item Three raters who are expert reviewers (researchers in HCI), working independently, rated a small sample of ideas using the initial definitions and anchors. 

    \item We met with the researchers as a group to discuss their sample ratings, focusing on areas of disagreement, and came to a shared agreement on the general definition of each criterion and what each of its scale value anchors meant. 
    
    \item Using these new definitions, we prompted GPT-4 to score a sample of ideas and to provide an explanation and justification for its assigned rating per criterion per idea. We then chose evaluation adjective and descriptive nouns from each explanation, and used these in a refined definition for a revised prompt. The definitions given in the prompt are: \textit{Relevance}: To what extent does the idea reflect how well the idea is connected with or appropriate for the objectives, requirements, or challenges of the problem statement? \textit{Innovation}: To what extent does the idea reflect how original and creative the idea is, breaking away from conventional or existing solutions to the problem statement? and \textit{Insightfulness}: To what extent does the idea reflect a profound and nuanced understanding of the problem statement?
    
    We repeated the process approximately three times for each scale until the anchors for each value were sufficiently differentiated.  Supplementary Information Figure 1 shows the prompt with the explanations for the various ratings per each criterion given to the GPT-4 evaluation engine. Figure~\ref{fig:engine} depicts an idea evaluation using the GPT-4 evaluation engine.

\end{enumerate}

\subsubsection{Implementation}
For this phase we chose GPT-4. At the time we conducted this experiment (June 2023), it has been available only for subscribers, and the researchers purchased a subscription. GPT-4 was chosen for the convergence phase over the free GPT-3 version due to its more advanced reasoning capabilities. We used the OpenAI API to write a Python program that uses the prompt to rate a set of ideas read in from a text file. The program outputs a CSV file with three ratings for each idea (for Relevance, Innovation, and Insightfulness), and a text file that contains GPT-4's justifications for those ratings. The user can indicate the number of times to repeat the process; each repetition will open a new GPT-4 context and produce a new set of ratings.

\section{User Study: Collaborative Group-AI Brainwriting Process}

We conducted a user study on the two stages of the collaborative Brainwriting process, the divergence stage and the convergence phase.
In the divergence stage, we integrated the use of GPT-3 into a Brainwriting session of an advanced undergraduate course on foundations of tangible interaction \cite{Ullmer22}. During a 70 minute session students followed the Brainwriting process described above. They first generated ideas independently, then worked with their team members to co-create ideas with GPT-3, and finally, chose ideas as a team to further develop through collaboration with GPT-3.

In the convergence stage, participants evaluated the quality of the ideas they generated throughout the session in terms of {\em relevance, innovation, and insightfulness} and chose a small final set of ideas. 

Following, we describe each part of the study in  detail.
\subsection{Divergence: The Collaborative Brainwriting session}
 
In February 2023, we conducted a 70-minutes Brainwriting session with 16 college students (0 men, ages 18-23) who were enrolled in an advanced undergraduate course on tangible interaction design. 
Considering the challenges interaction designers face when working with AI as a design material \cite{Dove17, Inie23, Yang20, Wang23, Subramonyam23}, this course aims to integrate co-creation and critical engagement with generative AI into its learning goals. Integrating the AI-augmented Brainwriting session into the course activities was thereby aligned with the course learning goals, among these: LG1) Apply a collaborative iterative process, which includes co-creation with AI and ML models for designing innovative tangible and embodied interfaces; LG2) Assess the capabilities and limits of prevalent AI technologies within the context of tangible interaction design; LG3) Implement a functional prototypes of a novel tangible or embodied interface using various technologies for data processing, sensing, and actuation. Develop AI intuition through experimental and creative exploration of AI technology for prototyping. The complete list of learning goals and course materials are available in the course website [link will be added in the camera ready version].  

 The students were divided into 5 project teams of 3-4 students each. The goal for the session was for students to start developing project ideas for a semester-long group project, which required them to \textit{“design a novel tangible user interface, which helps support the productivity, creativity, and well-being of people who work or study in mobile environments.”} Prior to the in-class Brainwriting session students were asked to read about Brainwriting \cite{Wilson2013} and about ChatGPT \cite{Roose2022, Perrigo2023}. 

After writing down their individual ideas on their team ConceptBoard, students used the OpenAI Playground GPT-3 to generate additional ideas using repetitive prompts.
We reminded students that ideating with GPT-3 might require multiple interactions in which they will need to refine their prompts and provided them with some examples for prompts used to generate similar tangible user interfaces (TUI) ideas. After adding the GPT-3 ideas to the board, we asked them to review, discuss, select, copy \& paste the best ideas to a side panel and start developing these ideas further with the help of GPT-3.

Table~\ref{table:numideas} shows the number of ideas generated by each team. The average word count of each Human-Generated idea is 16.5; the average word count of each GPT-3-Generated idea is 20.9.
In addition to submitting a link to their Conceptboard, students were asked to submit all their GPT-3 prompts.  

\begin{table}[!ht]
    \centering
    \captionof{table}{The number of ideas created per team: Human-Generated, GPT-3-Generated, Collaboratively-Generated, and total.}
    \begin{tabular}{|c|c|c|c|c|}
    \hline
     & \textbf{Human} &  \textbf{GPT-3} & \textbf{Collaborative} &\textbf{Total \# of ideas} \\\hline
  Team 1 & 20 & 4 & 2 & 26 \\ \hline
  Team 2 & 18 & 11 & 11 & 40 \\ \hline
  Team 3 & 17 & 2 & 0 & 19 \\ \hline
  Team 4 & 24 & 6 & 6 & 36 \\ \hline
  Team 5 & 18 & 6 & 3 & 27 \\ \hline
    \end{tabular}
    \label{table:numideas}
    \Description{The number of ideas created per team: Human-Generated, GPT-3-Generated, Collaboratively-Generated, and total}
\end{table}  

\subsection{Convergence: Ideas Evaluation and Selection } 
At the end of the session, the students were asked to rate the ideas: their own, GPT-3's and the collaborative ideas, as a means to narrow down the idea pool and engage in a selection process. The ideas were rated on a Likert scale along the three chosen evaluation criteria of relevance, innovation and insightfulness. Table~\ref{table:avgstd-1} shows the results of their self-ratings evaluation. The results show that students assign high levels of relevance, innovation, and insightfulness with mean scores of 4.75, 4.45, and 4.45, respectively to the ideas generated in their session. The distribution of scores exhibited a notable skewness, with 60\% of the questions attaining the maximum possible rating of 5 out of 5. 

\omitit{
\begin{table}[h!]
\centering
\caption{Average Self Ratings and Standard Deviations for Each Subcategory}
\begin{tabular}{lcccccc}
\toprule
 & \multicolumn{2}{c}{Relevance} & \multicolumn{2}{c}{Innovation} & \multicolumn{2}{c}{Insightful} \\
\cmidrule(lr){2-3} \cmidrule(lr){4-5} \cmidrule(lr){6-7}
Generated by & Avg & Std & Avg & Std & Avg & Std \\
\midrule
Human  & 4.81 & 0.40 & 4.31 & 0.70 & 4.37 & 0.61 \\
GPT-3    & 4.56 & 0.51 & 4.25 & 0.68 & 4.18 & 0.65 \\
Collab & 4.87 & 0.34 & 4.81 & 0.40 & 4.81 & 0.40 \\
\bottomrule
\end{tabular}
\label{table:avgstd}
\end{table}
}

\begin{table}[ht!]
\centering
\caption{Average Self Ratings and Standard Deviations for Each Evaluation Criterion}
\begin{tabular}{|l|c|c|c|c|c|c|}
\hline
 & \multicolumn{2}{c|}{Relevance} & \multicolumn{2}{c|}{Innovation} & \multicolumn{2}{c|}{Insightful} \\
\cline{2-7}
Generated by & Avg & Std & Avg & Std & Avg & Std \\
\hline
Human  & 4.81 & 0.40 & 4.31 & 0.70 & 4.37 & 0.61 \\
GPT-3    & 4.56 & 0.51 & 4.25 & 0.68 & 4.18 & 0.65 \\
Collab & 4.87 & 0.34 & 4.81 & 0.40 & 4.81 & 0.40 \\
\hline
\end{tabular}
\label{table:avgstd-1}
  \Description{Average Self Ratings and Standard Deviations for Each Subcategory}
\end{table}


After the session, each team chose an idea for their semester-long project. Table~\ref{table:finals} depicts the final ideas, and the source of the idea (human generated, LLM-generated, or combined).

Finally, we asked students about their experience Brainwriting with GPT-3 both immediately after the session, as well as again at the end of the semester. 

\section{Framework Evaluation}

The evaluation of the proposed collaborative group-AI Brainwriting framework consists of two parts. In the first, we explore through the use of qualitative and quantitative methods whether the use of LLMs in the divergence stage of group Brainwriting enhances the ideation process and its outcome (RQ1). To evaluate the \textit{quality} of the ideas, in addition to the participating students' self evaluation and to the ratings generated by the GPT-4 evaluation engine, three independent expert reviewers (HCI researchers) and six novice designers (HCI students) rated the quality of ideas on the same dimensions. Since the quality of ideas selected in the converge stage is impacted by the divergence of ideas generated \cite{coursey2019linking}, we evaluated divergence by examining the semantic distribution of ideas generated by humans and by GPT-3. We also identify the unique terms used in the different solution spaces.  We then explore, in the second part of the evaluation, how LLMs can be used to assist in idea evaluation during the convergence stage (RQ2). 

Here we describe the data and methods used in the evaluation of the proposed framework, followed by results organized by research question.

\subsection{Data and Methods}
We collected the following data: ideas generated by each team during the Brainwriting session; prompts used to interact with GPT-3; student responses to reflection questions; and novice designer ratings, expert ratings, and GPT-4 ratings.

We recruited 6 novice designers (students who completed an HCI course and were not enrolled in the same course in which we conducted the user study), as well as four expert reviewers who are active HCI researchers. Both novice and expert reviewers were asked to rate the set of ideas using the same three criteria definitions and scale value anchors given to the GPT-4 evaluation engine. The ideas given to the reviewers were arranged in a random order and there was no identifying information regarding the source of the idea (human or GPT-3). One expert reviewer provided evaluations for only a subset of ideas produced by student groups. In this document we report on data from the three expert reviewers who evaluated all of the ideas produced by students. 

We used thematic analysis \cite{braun2012thematic} to analyze the prompts used to interact with GPT-3 and the student reflection open responses. We first identified common keywords and tags among the responses, then aggregated these in order to extract broad themes and categories.

To examine the divergence of the ideas dataset (aggregated content of all 5 Conceptboards) we used the following methods and tools.  We first used the NLP toolkit spaCy to extract nouns and adjectives from the dataset. Also, we used spaCy and Gensim for topic modeling. We further use the Domain-based Latent Personal Analysis (LPA) method~\cite{mokryn2021domain}. LPA identifies the terms that most separate a document from a corpus. Using an Information-Theory approach, it creates a {\em signature} for each document, comprised of the terms that differ most in frequency in the document from their frequency in the corpus. These terms are corpus popular terms that are rare or missing in the document, and corpus rare terms that are frequent in the document. To create the signatures, each document is converted to a normalized term frequency vector, and the vectors are aggregated to create a corpus vector representation. LPA creates the signature per document by computing the symmetric per-element Kullback–Leibler Divergence (KLD)~\cite{kullback1951information}, also called relative entropy, between each document and the corpus. 
The relative entropy from distribution C to distribution D over sample space X is: 
\begin{equation}
\textit{KLD}(C||D)= -\sum_{x \in X} D(x)\log_2\left(\frac{C(x)}{D(x)}\right)
    \label{kld}
\end{equation}
LPA uses the symmetric KLD $\left(KLD(D||C)+KLD(C||D)\right)$ and pad document vectors with $\epsilon$-values for missing corpus terms. The corpus contains for each term that appeared in at least one of the documents its relative frequency. 
Here, as there are only two documents, one containing terms used by humans ($V_\mathcal{H}$) and the other the terms used by GPT-3 ($V_\mathcal{G}$), we perform the following. Each vector is expanded to contain all the terms in the $\cup(V_\mathcal{H},V_\mathcal{G}$) set, and missing terms are denoted as having zero frequency. The weight of each corpus term is computed as the average between the normalized term frequency in $V_\mathcal{H}$ and $V_\mathcal{G}$. 
Using Equation~\ref{kld} LPA finds for each document the terms that contributed most to the Relative Entropy of the terms that contributed most to the divergence of each of the normalized frequency vectors $V_\mathcal{H},V_\mathcal{G}$ from the corpus. 
 Term weights are assigned according to this contribution, with a corresponding sign. A positive sign indicates a rare corpus term that is overused in the document, and a negative sign indicates a corpus popular term that is underused or not used at all (missing) at the document. The set of terms with the highest absolute weight comprises the document's signature, each with its corresponding sign. 

Finally, statistical analysis was conducted using SPSS and Python. SPSS was used for hypothesis testing of agreement. GPT-4 was used for Semantic Analysis, and Python was used for descriptive analysis and LPA. 

\subsection{Results RQ1: Does the use of an LLM during the divergence stage of collaborative group Brainwriting enhance the idea generation process and its outcome?}

To answer RQ1 we examined both (a) student perceptions about the ideation process and (b) the outcome of the ideation process - the set of selected project ideas and their origin in terms of Human- and/or GPT-3-Generated ideas.
We then examined (c) the divergence of ideas through semantic analysis, and (d) the solution space explored with and without GPT-3 using LPA. Finally, we analyzed the (e) prompts used by students to interact with GPT-3. In the following, we describe the results.

\subsubsection{Students' Reflections} 
Since the user study was conducted within an educational setting, our evaluation of students' perceptions of this Group-AI Brainwriting framework also involved assessing their learning and critical engagement with AI. In a separate paper [currently under review for a different conference], we contextualized the use of this framework within a broader integration of generative AI into a tangible interaction course, and discussed students' reflections and learning. Here, we summarize student perceptions of the Group-AI Brainwriting process. Specifically, we analyze student responses to a question we asked immediately after the ideation session (Q1): \textit{"In what ways did using GPT-3 contribute to or hinder the ideation session?"} We also analyze their response to a question asked at the end of the semester (Q2): \textit{"Thinking back to your original ideation with GPT-3: to what extent do you feel like your collaboration with text-generative AI influenced the direction of your project?"}


\textit{Q1: In what ways did using GPT-3 contribute to or hinder the ideation session?}

All students responded to this question (n=16). Overall, we identified seven recurring themes: 3 themes describe positive contributions of GPT-3 to the ideation process, and 4 themes describe shortcomings of GPT-3. 50\% of students (8 out of 16) highlighted that GPT-3 offered them a unique or expanded viewpoint on the issue and its possible solutions. For example, one student shared that GPT-3 provided “ideas we had not offered or thought to offer on our own […] we were focused originally on one niche interpretation of the problem, and ultimately […] we got a more diverse set of possible products.” 44\% of students (7 out of 16) felt that GPT-3 significantly assisted them in generating ideas, in the words of one student: “adding in new ideas that we had not considered previously.” Some students  pointed out that their team(s) selected an idea for their concluding project that was initially suggested by GPT-3, one saying that “the model ultimately contributed the base idea we expanded upon with our own ideas to create the project pitch.” A smaller proportion of students (2 out of 16) mentioned that GPT-3 assisted them in articulating and communicating their own ideas. For example, one student wrote that “[GPT-3] helped us communicate our ideas better since it would reword our prompt.” 

 31\% of students (5 out of 16) pointed out that GPT-3 tends to be redundant and lacked creativity. For example, one student mentioned that “it didn’t come up with anything we didn’t.” Another student described their experience as though the AI was experiencing a “creative block,” they received similar results no matter how they reworded their prompt. 25\% of students (4 out of 16) reported challenges with crafting prompts and had to employ a trial and error approach to formulate prompts that produced high-quality responses from GPT-3. For example, one student shared that “there was a steep learning curve in understanding how to correctly prompt the model that hindered initial ideation.” One student expressed frustrations with the experience, describing how “it was pretty hard to get GPT-3 to output things the way we wanted it unless we used very specific language,” but added that their use of the tool was still helpful “as a way to kickstart our ideation and take the momentum into our own creativity.” Some students (2 out of 16) highlighted issues with the output being unrelated to the prompt or noted a lack of 'common sense' in understanding their request. One student shared their frustration with how GPT-3 would continually output “ideas that already existed,” such as Apple Watches.

\textit {Q2: Thinking back to your original ideation with GPT-3: to what extent do you feel like your collaboration with text-generative AI influenced the direction of your project?}

In response to this question, which was asked at the end of the semester, 50\% of the students (8 out of 16) indicated that using GPT-3 contributed to reshaping and enhancing their project by elaborating on their concepts, proposing new characteristics, and tackling particular challenges. In the words of one student: “The AI helped us reframe and refine our problem statements and questions so that may have been beneficial since we had to learn how to communicate with the AI. That alone made us more aware of the direction of our project since we had to refine the question on the spot in order for us to work with the AI. The AI also worked as a jumping off point for the team members to think of more organic, creative ideas.” Another student shared “I think that AI gave us many ideas that we could incorporate into [our project]. I think that collaboration with AI didn't necessarily generate an idea. However, with our specific idea in mind, we were able to utilize AI to think of more creative features.”

25\% (4 out of 16) said that GPT-3 had an impact on the direction of their project.  For example, one student wrote “It influenced the direction somewhat greatly - we already had the idea to make something that users could gather around, and use /after/ they had relocated while working remotely [...] but GPT-3 gave us the idea to make the [project] more community-oriented.” Another student shared “ChatGPT helped expand our brainstorming process and brought us ideas we hadn't thought of before, so we combined many into one as we decided on our project idea.” A few students described GPT-3 as a partner, assisting with particular tasks: “GPT-3 helped us with more specific information such as "how to alleviate motion sickness" and "what heartbeat threshold indicated an onset of motion sickness" that we did not inherently know. It was therefore helpful as a fourth teammate, but it could not replace any of us. So a nice companion, but not a substitute.” 

\subsubsection{Ideation outcomes}
The outcome of the human-LLM ideation process was a set of chosen ideas - each team chose one idea to explore in a semester long project. Table~\ref{table:finals} shows the chosen idea of each team, and describes the conception of each idea in terms of its human and/or GPT-3 origin. Overall, 3 out 5 chosen ideas were developed through merging a Human-Generated idea and a GPT-3-Generated idea. One idea was developed through merging multiple Human-Generated ideas and multiple GPT-3-Generated ideas.
Finally, one out of the 5 ideas is based solely on a GPT-3-Generated idea. 

\begin{table*}[!ht]
    \centering
    \caption{Brainwriting outcomes - a set of chosen ideas. For each team we describe the idea chosen for a project proposal and the enhancement type which contributed to its development. }
     \begin{tabular}{|l|p{4.2cm}|p{2.3cm}|p{4.2cm}|}
    \hline
         & Chosen idea & Enhancement & Description \\ \hline
        Team 1 & An interactive public display that allows local users to "pin" their preferred working spots; travelling workers coming into town can check out the interactive map via their mobile phone & Combined human \& LLM & Inspired by the combination of a Human-Generated idea, a platform for rating work spaces, with a GPT-3 Generated idea, an interactive public display \\ \hline
        Team 2 & Posture pillow that keeps track of posture patterns and reminds user to change their position or take a break & LLM & Inspired by a GPT-3-Generated idea for a smart pillow that can detect posture \\ \hline
       Team 3 & Portable desk for commuter students with stability and motion sickness-alleviating features and a built-in wifi hotspot & Combined human \& LLM & Not submitted with original workshop idea set, but submitted with project proposal as a combination of a Human-Generated idea ("portable desk that is stable on bumpy rides") and a GPT-3-Generated idea ("installing a wireless router or access point inside a portable desk") \\ \hline
       Team 4 & A plushie/stress ball keychain that users can hold onto; releases aromatherapy and also communicates with user holding another one to either feel their heartbeat or the same squeezing sensation & Combined human \& LLM  & Inspired by combining a number of Human-Generated and GPT-3-Generated ideas having to do with aromatherapy for stress and paired devices that transmit the users' pulse. Unlike the other teams, this team combined several ideas together \\ \hline
      Team  5 & Sleeping eye mask that changes temperature based on where you are in your journey and vibrates to wake you before your stop & Combined human \& LLM  & Inspired by the combination of a Human-Generated idea, a wearable to notify the user when their public transport stop is near, with a GPT-3-Generated idea, a temperature-controlled sleep mask \\ \hline
    \end{tabular}
    \label{table:finals}
    \Description{Brainwriting outcomes - a set of chosen ideas. For each team we describe the idea chosen for a project proposal and the enhancement type which contributed to its development.}
\end{table*}


\subsubsection{Exploring the Human and LLM solution spaces} 
To explore the divergence of ideas and the solution space explored with and without LLMs, we evaluated the  semantic distribution of ideas generated by humans and by GPT-3, and the  terms used in the different solution spaces using LPA. 

Evaluating the semantics of different idea spaces allows us to explore potential conceptual differences between the human and AI idea spaces. If these concept spaces, as determined by our methods, show substantial overlap, it would suggest that in this experiment, the AI did not significantly augment the human creative thought process from a conceptual aspect. For the evaluation, we compared a semantic clustering over the terms used in these spaces, and then evaluated the differences in the terminology. A difference in terminology can be semantic or more substantial. A substantial difference, characterized by overused, underused, or entirely absent terms in a solution space, offers deeper insights into the variances that may exist between human and AI-generated ideas.


\paragraph{Semantic clustering analysis. }
To discuss the semantic clustering analysis, the following terminology is used.  The set of Human-Generated ideas as $\mathcal{H}$, and the set of GPT-3-Generated idea as $\mathcal{G}$. The semantic analysis was done by generating semantic clusters of the ideas in both sets, $\mathcal{H}$ and $\mathcal{G}$.   The semantic analysis of $\mathcal{H}$ yielded 20 clusters, and of $\mathcal{G}$ 21 clusters. There were 12 similar clusters that  contained shared terms. For example, in both sets the cluster \text{Digital Devices \& Hardware} contained the terms <\textit{computer, monitor, laptop, smartphone (and phone), tablet}>, and the cluster \text{Health \& Wellness} contained the terms <\textit{sleep, meditation, stress, nausea, heartbeat, pulse}>. 
The semantic clustering of $\mathcal{H}$ contained the following unique clusters and terms: Vehicle-related terms <\textit{bus, commute, train}>, Personal clothing <\textit{jacket, sweater}>, Food and beverages <\textit{dining, water}>, Learning \& information <\textit{study, academy, library}>, and Games \& entertainment <\textit{Pokemon, music, leisure}>. The semantic clustering of $\mathcal{G}$ contained \text{Screen and display elements} <\textit{background, settings}>, Interactivity and controls <\textit{buttons, dials, gestures}>, Specific measurements <\textit{cm, diameter, intensity}>, Visual \& design elements <\textit{shapes, signs}>, and Specific work-related terms <\textit{brainstorming, distractions}>. The full list of clusters and their corresponding terms can be found in the Supplementary Information. 

Overall, while many of the semantic clusters were similar, the differences seem to relate to the level of detailing of the concepts.  The concepts found only in $\mathcal{H}$ tended to be more abstract or alluded to objects in a generalized manner, while concepts found in $\mathcal{G}$ were more concrete or pertained to specific details of objects or their description, such as their measurements. 

\paragraph{LPA of the terminology used in the two solution spaces}
\begin{figure*}[!ht]
    \centering
    \begin{subfigure}{\textwidth}
       \centering     
\includegraphics[width=\linewidth]{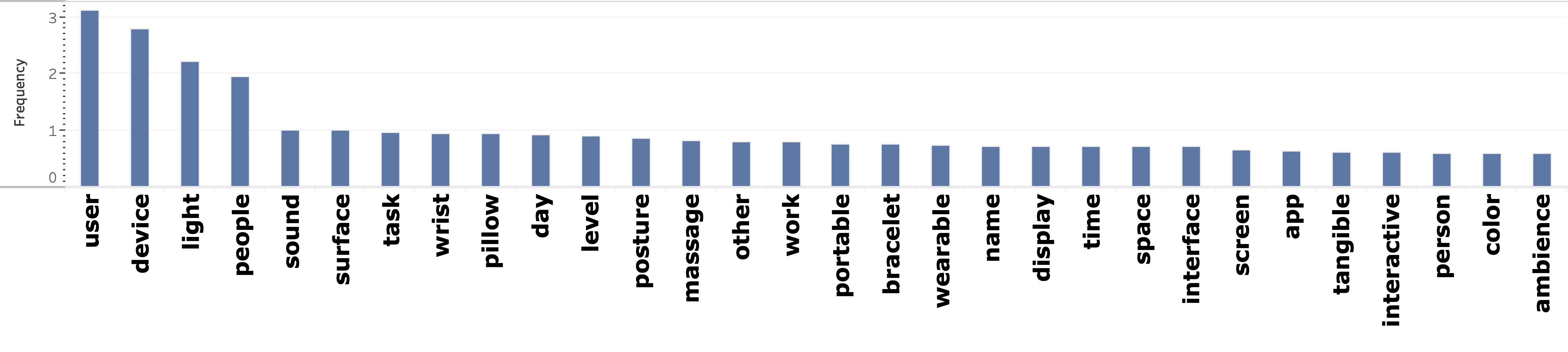}
        \caption{The top used terms in the joint vocabulary created by LPA for the terms used by humans and GPT-3 in their ideas.}
        \label{fig:dvr}
    \end{subfigure} 
    \begin{subfigure}{\textwidth}
       \centering
\includegraphics[width=0.95\linewidth]{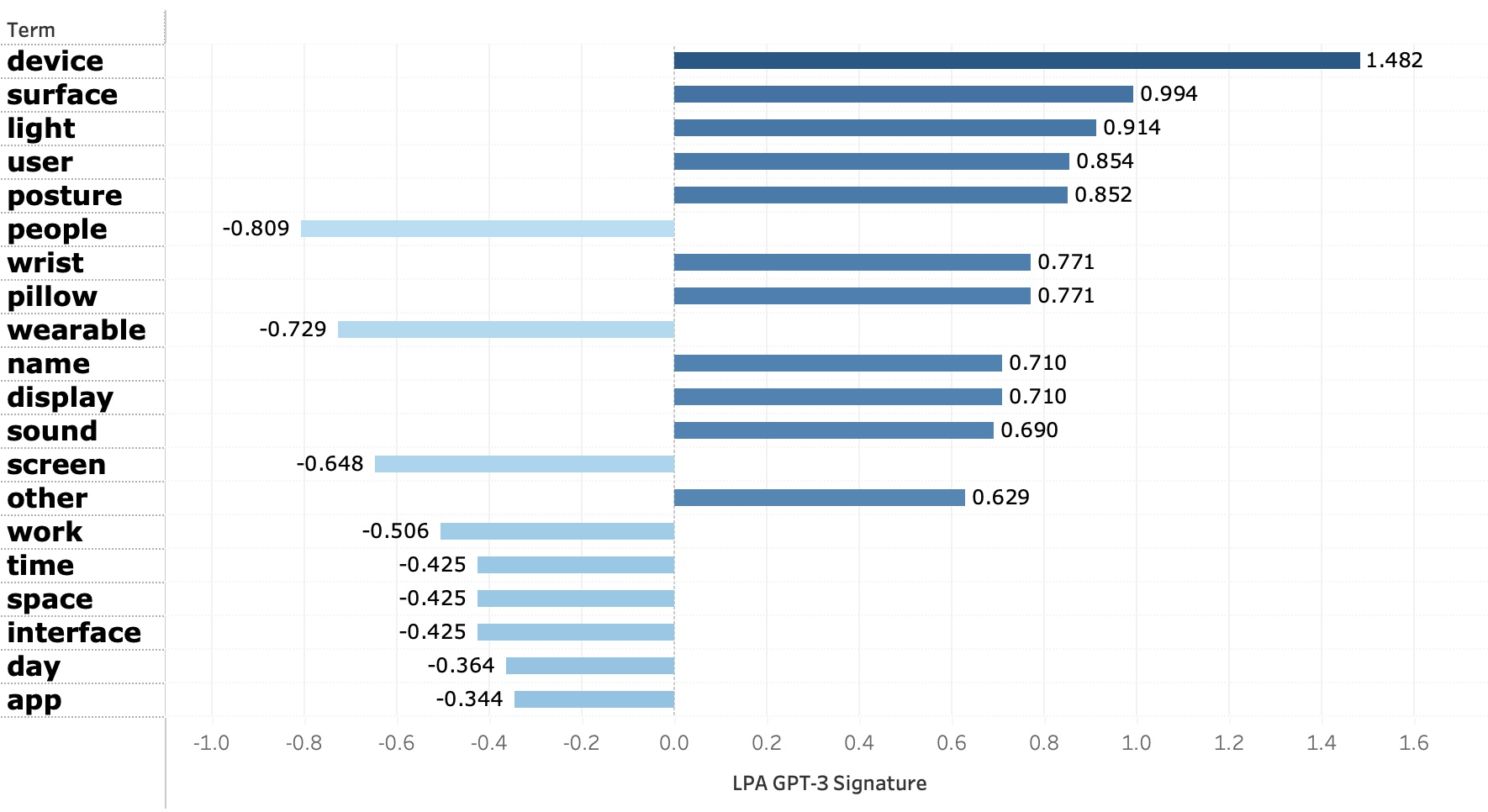}
        \caption{GPT-3 LPA signature, denoting terms that are overused in ideas compared to humans, and terms that are underused or missing, when compared with humans. The corresponding weight of each term denotes the difference in the percentage of times it is used in GPT-3's ideas  compared to the average usage across the ideas generated by either humans or GPT-3.\\}
        \label{fig:gpt}

    \end{subfigure}
    \caption{Identifying biases in LLM-generated ideas. (a) introduces the top terms used in all ideas generated either by humans or by GPT-3, as calculated using the Latent Personal Analysis (LPA) method. (b) depicts GPT-3's LPA signature, denoting its unique use of terms when compared to the shared vocabulary, either underused or overused.}
    \label{fig:lpa}
    \Description{Identifying biases in LLM-generated ideas. Lower panel (a) introduces the top terms used in all ideas generated either by humans or by GPT-3, as calculated using the Latent Personal Analysis (LPA) method. Upper panel (b) depicts GPT-3's LPA signature, denoting its unique use of terms when compared to the shared vocabulary, either underused or overused.}
\end{figure*} 

Here, we examine the differences in noun terms used within the solution spaces of LLMs and human-generated ideas. Variations in noun term usage can reveal conceptual or thematic differences, highlight the level of detail and depth in the ideas, indicate their specificity and breadth, and may also suggest the context to which the idea pertains. LPA identifies the main differences between the two corresponding noun terms distributions. 

LPA analysis of the terms used either by humans or GPT-3 reveals a difference. Figure~\ref{fig:lpa} shows the results of the analysis. The ten most prevalent terms used in ideas by either humans or GPT-3 (normalized to account for the different number of ideas in each group) were \textit{user, device, light, people, sound, surface, task, wrist, pillow, day}, depicted in Figure~\ref{fig:dvr}. 
However, there were some notable differences, as can be seen from GPT-3's LPA signature, depicted in Figure~\ref{fig:gpt}. For example, while ideas created by humans referred to \textit{people}, GPT-3 kept using the term \textit{users}.  The term \textit{device} was prevalent in GPT-3's ideas, while hardly used by humans. Other GPT-3's prevalent terms were \textit{surface, light, posture, wrist} that were hardly used by humans. 
On the other hand, GPT-3 did not refer to terms that were commonly used in human ideas, such as  \textit{wearable, screen, work, time, space, interface, day, app}.

\subsubsection{Prompt analysis}

To get further insight into the differences between Human-Generated and GPT-3-Generated ideas, we analyzed the prompts used by students to generate new ideas and iterate on existing ones, and identified a few distinct approaches.  Typically, students used one of two approaches to initiate their interaction with GPT-3: 1) broad-area prompts, or 2) solution-specific prompts. 

Broad-area prompts involved giving GPT-3 an open-ended request for ideas related to the problem statement. For example, one team began their interaction with the prompt, “Tell me a list of ideas for tangible interfaces that support productivity and creativity that doesn't exist yet ”. Solution-specific prompts entailed asking for a solution for a concrete problem. For example, “Tell me ways to stabilize a portable desk when on a bus”; 

When students decided to focus on a particular idea, they applied two different approaches to expand on their idea: 1) usage-focused follow-up prompts, and 2) detail-focused follow-up prompts. A usage-focused prompt asked GPT-3 to expand on the ways and context users would use their proposed solution. For example, one team asked “How can this device be utilized without Wendy having to change the settings?” A detail-focused prompt, on the other hand, asked GPT-3 to expand on the features and capabilities of a specific idea. For example, “Tell me a list of functionalities that a smart light can do to make you more productive and creative.”

Student teams combined these approaches during the ideation session. 

\subsubsection{Summary of findings for RQ1}

After the session, 50\% of students perceived GPT-3 as helpful because it provided a unique or expanded perspective on the problem statement and its possible solutions. 44\% shared that it  significantly assisted them in generating new ideas. At the end of the semester, 50\% of the students mentioned that GPT-3 contributed to reshaping and enhancing their project by elaborating on their concepts, proposing new characteristics, and tackling particular challenges. 31\% of students pointed out that GPT-3 tends to be redundant and lacked creativity. 

The ideas chosen by each group for their final project were mostly created by combining an idea generated by team members and an idea suggested or enhanced by the LLM. In one case (Team 2), the chosen idea was directly inspired by an idea generated by GPT-3. 

Semantic clustering analysis of Human- and GPT-3-Generated ideas indicates that 
humans tended to allude to abstract concepts and refer to objects in a general way, while the ideas generated by GPT-3 were more concrete. 
The solution space, denoted by the different vocabulary used in ideas generated by humans and GPT-3,  is consistent with these findings. For example, the term ``device'' appears almost exclusively in GPT-3-Generated ideas, which often also reference their ``users''. In Human-Generated ideas, the reference is to ``people'', and the term ``wearable'' appears only in human ideas. Humans tend also to refer more to ``space'' and ``time'', while GPT-3 referred more to ``surface'' and ``light''. 

The prompt analysis reveals that students combined approaches when interacting with GPT-3, typically starting with a broad request for ideas, then requesting solutions for a concrete problem, or asking for additional details regrading the usage, features, and/or capabilities of a specific idea.  These results explain, to some extent, the higher level of details we found in GPT-3-Generated ideas.

\subsection{RQ2: How can LLMs assist to evaluate ideas during the convergence stage of a collaborative group Brainwriting process?}

We assess here the feasibility of using an LLM to assist in idea evaluation in the convergence phase. These idea evaluations were not part of the User Study and were conducted after the student deadline for choosing the final ideas. To evaluate how LLMs can help in the convergence phase, in which all ideas are evaluated and a few are selected, we assess here: (a) whether LLMs' evaluations are consistent, and (b) how they compare with evaluations made by experts and novices. Our goal here is to assess whether LLMs can be used to filter out ideas reliably.

All ideas created during the Brainwriting process: Human-Generated, GPT-3-Generated, and Collaboratively-Generated, were evaluated by 3 Experts, 6 Novices, and the GPT-4 evaluation engine. All evaluations used the same 1 to 5 Likert Scale for Relevance, Innovation, and Insightfulness. Both Novice and Expert reviewers were given the same criteria definition and scale value anchors given to the GPT-4 evaluation engine. The ideas given to the reviewers were arranged in a random order and there was no identifying information regarding the source of the idea (human or GPT-3). The GPT-4 engine was prompted to repeat each evaluation 30 times (29 rounds were completed successfully), each evaluation conducted in a new context. 

\subsubsection{Consistency of the GPT-4 evaluation engine.} 
First, we assess the internal consistency of the 29 GPT-4 evaluations for the ideas on the three criteria of Relevance, Innovation, and insightfulness.  To evaluate consistency we treat the evaluations as questionnaire items and analyze them with Fleiss' Kappa coefficients to evaluate rater agreement. Our analysis shows a moderate level of consistency in GPT-4's performance, with all Fleiss' Kappa values surpassing the 0.4 threshold. The specific Fleiss' Kappa values for the different criteria were the following. Relevance: 0.42, Innovation: 0.40, and Insightfulness: 0.49. Thus, GPT-4 evaluations can be seen as consistent across the three criteria. 

\subsubsection{Comparative Analysis of GPT-4's Evaluations Against Novice and Expert Human Evaluators. } 
\label{sec:compare}
We compare the ratings given GPT-4 to those given by novices and experts to the 148 ideas generated by either humans, GPT-3, or in collaboration. The ratings were given to each idea for each of the three criteria: Relevance, Innovation, and Insightfulness. To compare the GPT-4 evaluations to human raters, we conducted the following steps:  (a) compared the given rating distributions, (b) compared evaluations for the top and bottom ideas as ranked by the experts' ratings; (c) computed the Pearson correlation between GPT-4 ranking of ideas and the experts' ranking; (d) compared the ratings given by GPT-4, novices, and experts, across the three criteria, to the ideas that were chosen by the teams as their final ideas. 

Unlike GPT-4, Expert and Novice evaluators had diverging opinions and medium to low internal consistency across the three criteria. A Shapiro-Wilk Test on the raw rating distribution of Experts evaluations found that the null hypothesis of a normal distribution is rejected with a p-value << 0.001 for the ratings of all three criteria: Relevance, Innovation, and Insightfulness. Similarly, the Shapiro-Wilk Test on the raw rating distribution of Novice evaluations found that the null hypothesis of a normal distribution is rejected with a p-value of << 0.01 for all three criteria. 

\begin{figure}
\includegraphics[width=0.9\linewidth]{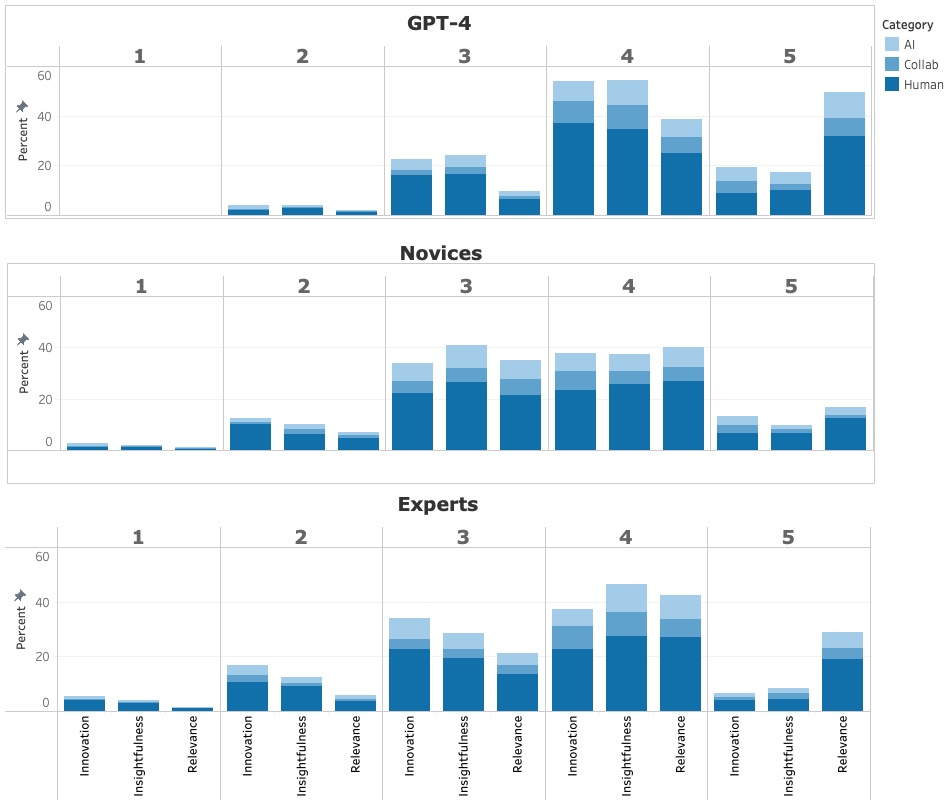}
  \caption{The Distribution of ratings on a 1 to 5 Likert scale given to ideas generated in the Brainwriting process. Ideas were generated by either humans, GPT-3, or as a collaboration. Every idea was assessed based on three  criteria: its relevance, depth of insight, and level of innovation.  All 148 ideas were rated by Experts, Novices, and the GPT-4 rating engine. 
The lower panel depicts the distribution of ratings given by Experts to ideas in each of the criteria.  The middle panel depicts ratings given by Novices, and the upper panel the rates given by GPT-4.  }
  \label{fig:desc-raters}
    \Description{ The Distribution of ratings on a 1 to 5 Likert scale given to ideas generated in the Brainwriting process. Ideas were generated by either humans, GPT-3, or as a collaboration. Every idea was assessed based on the three criteria: its relevance, the depth of insight it demonstrated, and its level of innovation. All 148 ideas were rated by Experts, Novices and the GPT-4 rating engine. Lower panel depicts the distribution of ratings given by Experts to ideas in each of the criteria. Middle panel depicts ratings given by Novices, and upper panel the rates given by GPT-4.}
\end{figure}

(a) First, we compare the ratings distributions across the evaluator groups. Figure~\ref{fig:desc-raters} depicts the distribution of ratings on a Likert scale of 1 to 5 given by Experts (lower panel), Novices (middle panel), and GPT-4 (upper panel) for the 148 ideas across the three criteria.  For each idea and criterion, the rating was calculated as the average of the ratings given by the corresponding rater group, either Experts, Novices, or GPT-4, to that idea.  
The ratings distributions demonstrate that the Experts were more critical than the Novices and that GPT-4 gives relatively high ratings to ideas. GPT-4 gave much more ratings of 5 than novices and experts and much less ratings of 2 and 1. Specifically, it gave a lower rating of 1 to only one idea, for its Insightfulness.  GPT-4 gave an average rating of 4.19 for relevance, 3.72 for innovation, and 3.68 for insightfulness.

Clearly, there is no agreement between either of the groups, and hence also not with that of GPT-4.  We then continue to examine the similarity in ranking of ideas, and the ratings given to the final ideas as chosen by the teams. 

(b) We created a ranking of the ideas for each rater group, Experts, Novices, and GPT-4.  The ranking of the ideas was computed as follows. For each rater group, the rating of an idea by that rater group was computed by averaging the ratings given by the group members for each of the criteria and then by summing these values. For example, in the case of the Expert rater group, the average rating given by the three experts to each of the criteria Relevance, Innovation, and Insightfulness was computed, and the idea's final rating was computed as the sum of these three average values. per each criteria and summing it. Thus,  an idea with an Experts average rating of 4 for relevance, 2.75 for innovation, and 2.375 for insightfulness received an aggregated rating of 
$9.125$, and was ranked $24$ out of 148 ideas. 

From the Expert ranked idea list, we chose the four highest and lowest-ranked ideas and compared their ranking to their ranking on the GPT-4 ranked idea list.  On the Expert ranking list, the top four received ratings of 13, 12.5, 12.5, 12.5. The lowest-ranking ideas received ratings of 5.75, 5.33, 5.25, 5.25. 
Of the four ideas ranked highest by the Experts, one was also in the second place on GPT-4 list, and the rest were in the top half of the list. 
Out of the four ideas ranked lowest by the experts, three were in the bottom 6 places on GPT-4 ranked list. The fourth ranked lowest idea by the Experts, was ranked in the middle of the list by GPT-4. 

Comparing the top and bottom four between experts and novices, we found that out of the experts' four top rated ideas two were also rated at the top by novices. The two other ideas were not at the top of the novices' list. There was no agreement at the bottom part of the ranked list, as all ideas that were rated lowest by the experts appeared in the lower quarter, however not in the bottom of the novices' list. 
When comparing the novices' and GPT-4's top and bottom elements, we find that there is a high agreement.

(c) To quantify the relationship between the rankings provided by different groups, we computed Pearson correlation coefficients. The comparison yielded a coefficient of 0.556 between expert and GPT-4 ratings, 0.547 between novice and GPT-4 ratings, and 0.602 between expert and novice ratings. These results indicate a moderate positive linear relationship among the three ranked lists.

Thus, we can conclude that overall, GPT-4's ranking of ideas is generally in agreement with the Experts' and novices' rankings. 

\begin{table*}[!ht]
    \centering
    \caption{Comparison of the evaluations of experts, novices, and GPT-4 for the chosen final project ideas of each team, as described in Table~\ref{table:finals}}
\begin{tabular}{|l|ll|r|r|r|r|r|r|}
\hline
Rater  & \multicolumn{2}{c|}{Criterion}
       & \multicolumn{1}{l|}{Team 1} 
       & \multicolumn{1}{l|}{Team 2} 
       & \multicolumn{1}{l|}{Team 3} 
       & \multicolumn{1}{l|}{Team 4} 
       & \multicolumn{1}{l|}{Team 5} 
       & Average over all ideas \\
\hline
       & Relevance      & Avg   & 3.75 & 4.25 & 4.00 & 3.67 & 3.00 & 3.57\\
       &                & Stdev & 0.96 & 0.50 & 1.00 & 0.58 & 1.83 & 1.10\\
        \cline{2-9}
Expert & Innovation     & Avg   & 3.00 & 3.25 & 3.33 & 3.00 & 2.00 & 2.79\\
       &                & Stdev & 1.41 & 0.96 & 2.08 & 1.00 & 0.82 & 1.10\\
        \cline{2-9}
       & Insightfulness & Avg   & 3.00 & 3.25 & 3.67 & 3.67 & 2.25 & 3.01\\
       &                & Stdev & 1.15 & 1.26 & 1.53 & 0.58 & 1.26 & 1.11\\
\hline
       & Relevance      & Avg   & 3.67 & 3.50 & 4.17 & 3.17 & 3.33 & 3.38\\
       &                & Stdev & 0.52 & 0.55 & 0.75 & 0.75 & 0.82 & 0.95\\
        \cline{2-9}
Novice & Innovation     & Avg   & 2.83 & 3.67 & 3.50 & 3.83 & 3.83 & 3.11\\
       &                & Stdev & 0.98 & 0.52 & 1.05 & 0.98 & 0.75 & 1.07\\
        \cline{2-9}
       & Insightfulness & Avg   & 3.50 & 3.67 & 3.50 & 3.33 & 3.17 & 3.13\\
       &                & Stdev & 0.55 & 0.52 & 0.84 & 1.03 & 0.98 & 0.96\\
\hline
       & Relevance      & Avg   & 4.80 & 4.73 & 4.52 & 4.03 & 4.57 & 4.19\\
       &                & Stdev & 0.41 & 0.45 & 0.51 & 0.32 & 0.50 & 0.82\\
        \cline{2-9}
GPT-4  & Innovation     & Avg   & 3.77 & 3.90 & 3.52 & 4.57 & 4.93 & 3.72\\
       &                & Stdev & 0.43 & 0.31 & 0.51 & 0.50 & 0.25 & 0.80\\
        \cline{2-9}
       & Insightfulness & Avg   & 3.87 & 4.27 & 3.93 & 3.87 & 4.33 & 3.68\\
       &                & Stdev & 0.43 & 0.69 & 0.53 & 0.43 & 0.48 & 0.80\\              
\hline
\end{tabular}
    \label{tbl:chosen}
    \Description{Comparison of the evaluations of experts, novices, and GPT-4 for the chosen final project ideas of each team}
\end{table*}

(d) Lastly, we examine the evaluations given by the GPT-4 evaluation engine to the ideas that were ultimately chosen by student teams, and compare these to the expert and novices corresponding ratings. Table~\ref{tbl:chosen} summarizes the the ratings of the final ideas chosen by teams. 
For the majority of instances, all rater groups, namely experts, novices, and the GPT-4 evaluation engine, assigned higher ratings to the final selected ideas compared to the average rating they assigned to all ideas. 

We have shown that both the expert and the novice raters had diverging opinions on many of the ideas. While the majority of the final project ideas received a higher rating than the average idea from the experts (but team 5's idea), their evaluations of the ideas along the three criteria differ substantially, as  reflected in the relatively high standard deviation values. Similar disagreement, although to a lesser degree, exists also among the novice raters' evaluations. 

Among the evaluations of the teams chosen ideas, the largest disagreement between the rater groups exists between the experts and GPT-4 for team 5 chosen idea. The largest difference exists for the evaluation of the Innovation of the idea, receiving a low average score of 2.00 by the experts, compared with an average rating of 4.93 from GPT-4. Interestingly, this idea received the highest rating for Innovation from the novice raters among the chosen ideas. 

Overall, our analysis shows that GPT-4 evaluation engine did not rate below the average the ideas that were chosen as final by the ideas.

\subsubsection{Summary of findings for RQ2}

The GPT-4 evaluation engine gave high ratings to all of the ideas that were ultimately chosen by student teams as can be seen in Table~\ref{tbl:chosen}. We further observed a robust level of internal consistency among the ratings generated by the GPT-4 engine, as evidenced by elevated values of Fleiss' Kappa exceeding 0.4 across all three criteria: Relevance, Innovation, Insightfulness. Unlike GPT-4, Expert and Novice evaluators had diverging opinions, and medium to low internal consistency across the three criteria. The distributions of evaluations reveal that Experts were more critical than Novices, and that GPT-4 gives relatively high ratings to ideas. 

We evaluated the alignment of idea rankings between experts, novices, and GPT-4. A notable correlation was observed, especially between the highest and lowest-rated ideas. Top ideas as rated by experts were generally also favored by GPT-4, with a similar pattern evident in the novice evaluations. The Pearson correlation coefficients – 0.556 between experts and GPT-4, 0.547 between novices and GPT-4, and 0.602 between experts and novices – suggested a moderate positive linear relationship among the three groups' rankings. This consistency across human and AI evaluations highlights GPT-4's potential as a viable tool for preliminary idea filtering, aligning closely with human judgment in identifying high-quality ideas.

 The fact that none of the chosen ideas received low ratings by GPT-4 is encouraging - it means that, if GPT-4 had been used to provide feedback for teams during the ideation process, it would not have filtered out ideas that were considered to be good by the teams. At the same time, it also appears that, had GPT-4 been used to provide feedback during the ideation process, teams could have safely discarded ideas that were rated low by GPT-4. After all, none of the ideas that were rated low by GPT-4 were ultimately chosen. (Note that we used GPT-4 to evaluate ideas only after the ideation sessions were completed, so these evaluations were not available to teams.)
\section{Discussion}

In this paper we propose a framework for collaborative group-AI Brainwriting and study two dimensions of such integration. First, we study the use of an LLM for enhancing the idea generation process. Second, we explore the use of an LLM for evaluating ideas during the convergence phase, in which three criteria of the ideas are evaluated: their relevance to  the problem statement, the originality and creativity of the idea, i.e., how innovative it is, and the extent to which the idea reflects a profound and nuanced understanding of the problem statement, which we refer to as the insightfulness of the idea. 
We conduct a user study that uses the framework for an idea generation process as part of a college-level interaction design course, and conduct a set of evaluations of the process, its outcomes, and the potential use of an LLM for the evaluation process.

Here we discuss our findings, focusing on addressing the two research questions we introduced in the introduction. We then discuss implications for HCI education and practice.

\subsection{Discussion of results for RQ1: Does the use of an LLM during the divergence stage of collaborative group Brainwriting enhance the idea generation process and its outcome?}

In their reflections, 50\% of the students found the use of GPT-3 helpful in providing unique or expanded perspective on the problem statement and its possible solutions. Findings from our semantic and LPA analyses of the idea space, indicate that indeed GPT-3 contributed both ideas that were somewhat different from those generated by humans, as well as included more technical and usage details. These findings indicate that integrating an LLM into the Brainwriting ideation process could provide support for both divergent thinking - producing a wide range of different ideas, and convergent thinking - incremental, step-by-step development of the details of a solution \cite{TverskyCreativity}. Indeed, the set of chosen ideas (see Table~\ref{table:finals}) illustrates that GPT-3 provided enhancements to the ideation process - all 5 teams chose project ideas that either combine GPT-3-Generated ideas with Human-Generated ideas, or are based on a GPT-3-Generated idea.


However, in our study, about 30\% of the students pointed out that GPT-3 tends to be redundant and lacks creativity. How can we increase the novelty and creativity of the ideas contributed by an LLM to a collaborative group-AI ideation process?  One possibility is through prompt engineering. In our study, students prompt the GPT-3 model directly, but integrating an LLM model into a custom interface, which implements back-end prompt engineering could potentially cause the LLM to provide better assistance for users during ideation. Several tools demonstrate the use of back-end prompt engineering within the context of education (e.g.\cite{Han2023RECIPEHT, Kumar_Musabirov_Reza_Shi_Kuzminykh_Williams_Liut_2023b}) and decision making (e.g. \cite{Park_Min_Ma_Kim_2023}).



Applying this approach, we can help users to utilize prompts that challenge conventional molds. One direction is through connecting seemingly unrelated concepts in a way that invokes conceptual blending - a cognitive process in which distinct ideas are combined to create a new, unique concept~\cite{FAUCONNIER1998133}. Wang and colleagues have demonstrated the feasibility of this approach with a system that automatically suggests conceptual blends~\cite{Wang_Popblend}. Another possibility is to adopt an approach similar to "Six Thinking Hats"~\cite{Bono1999}, where different prompts are constructed, each defining a different persona for the LLM and hence leading to ideas that are provided in different style and represent different perspectives. Yet another approach might be an adaptation of the process proposed by Kahneman and colleagues to reduce noise in decision making -- the authors propose that decision maker teams approach a problem by separating it into well-defined and separate focus areas \cite{kahneman2021noise}. For us, this could mean crafting different prompts that aim to elicit ideas about different aspect of the question at hand, e.g. a technological implementation, or an issue of aesthetics.  



\subsection{Discussion of results for RQ2: How can LLMs assist to evaluate ideas during the convergence stage of a collaborative group Brainwriting process?}

The GPT-4 evaluation engine gave relatively high ratings to all of the ideas that were ultimately chosen by student teams, see  Table~\ref{table:finals}. The fact that none of the chosen ideas received low ratings by GPT-4 is encouraging - it indicates that, if GPT-4 had been used to provide feedback for teams during the ideation process, it would not have filtered out ideas that were considered to be good by the teams. 

At the same time, based on the moderate positive linear relationship between Expert and GPT-4 engine review scores, it also appears that, had GPT-4 been used to provide feedback during the ideation process, teams could have safely discarded ideas that were rated low by GPT-4. After all, none of the ideas that were rated low by GPT-4 were ultimately chosen, and none of the ideas that were rated low by Experts were rated high by GPT-4. 



A final note on how LLMs can be used in supporting idea evaluation relates to the statistical terms of noise and bias~\cite{kahneman2021noise}. 
Statistically, we saw that GPT-4 made consistent decisions as we asked it to evaluate each idea 29 times; thus, the noise in GPT-4 decisions was low. However, on average, GPT-4 and Expert evaluations differed from each other, representing a statistical bias. It is clear that this statistical observation in our data can translate into future versions of an LLM system that attempts to support ideation but provides feedback with harmful biases.

\subsection{Implications for HCI education and practice}

While generative AI have created new opportunities for supporting designers~\cite{hbr2022GenerativeAI, Olsson21}, the structured integration of AI into design courses remains challenging~\cite{Flechtner23}. In this paper we contribute a practical framework for collaborative group-AI Brainwriting that could be applied in HCI education and practice. We evaluated this framework with college students as part of their project work in a tangible interaction design course. The integration of co-creation processes with AI was aligned with the learning goals for the course, which aims to address some of the challenges that designers face when working with AI as design material~\cite{Dove17, Inie23, Yang20, Wang23, Subramonyam23}. Here we discuss the implications of our findings for HCI education and practice.

\subsubsection{Expanding Ideas}
Our findings demonstrate that integrating co-creation processes with AI into the ideation process of novice designers, could enhance the divergence stage where a wider range of different ideas is explored. 

From our experience teaching tangible and embodied interaction design over the years [hidden for anonymity], students or novice designers who are new to TEI often limit their early ideation to traditional forms of interaction such as mobile phone apps and screen-based wearables. Results from our brainwriting activity indicate that using an LLM during ideation helped students to expand their ideas, and to consider different approaches (see Figure~\ref{fig:lpa}). While the creativity exhibited by GPT-3 itself was sometimes limited when prompted for producing new ideas, when it was prompted to expand on specific students’ ideas, it often provided new modalities and suggested novel features that diverged from traditional graphical user interfaces (see section~\ref{sec:compare}).

\subsubsection{Prompt Engineering}
The comments made by the students in our study make it clear that sometimes they struggled with creating effective prompts for GPT-3. This is an important issue, since our goal is to support ideation for teams with diverse levels of experience working with LLMs, not only professionals with training in the usage of the latest LLM technologies. While back-end prompting is one approach to address this challenge, it is clear that novice designers also require instruction on constructing effective prompts. It is thereby important to develop training materials for interaction designers in best practices of prompt engineering and to encourage them to consider how best to provide domain, task, and interaction style specific keywords with their prompts.

\subsubsection{Increased Creativity through Shifting Attention}
Tversky and Chou suggest that shifting attention between different problems fosters divergent thought and enhance creativity \cite{TverskyCreativity}. Future variation of our proposed framework for collaborative group-AI Brainwriting, could shift the group attention so that an LLM is prompted multiple times, where each prompt is focused on a different problem or aspect of the problem. Future research should explore such strategies for increasing the creativity of LLM-generated ideas. 

\subsubsection{Limitations of Non-Human Agents}
The proposed group-AI Brainwriting process could be considered within the realm of more-than-human, post-humanist interaction design methods~\cite{Giaccardi20, Homewood21,Wakkary2021, wakkary2020nomadic} where agency is distributed among humans and non-human agents such as LLMs. When applying such methods it is important to remember that AI-based non-human agents are trained upon and import "traditional, humanist forms of logic and language"~\cite{vanDijk2020}. Thus, co-creation ideation processes might yield ideas that embody and amplify human social biases. While we did not identify specific social biases in the ideas produced by the proposed group-AI collaboration in response to the given problem statement, future work should probe for ideas that contain bias regarding specific groups or concepts. Future work could also develop methods of filtering out ideas that contain such bias.

\subsubsection{Evaluating Ideas}
In our study, we used the GPT-4 evaluation engine only after the ideation sessions were completed, so these evaluations were not available to teams. As we continue to work towards providing such LLM-generated evaluations to users, there are several issues to consider. First, such use of LLMs falls into the trend identified by Janssen et al.~\cite{janssen2019history} that automation is increasingly being used by users with varied levels of expertise in using automated and AI-powered tools. LLM-generated feedback needs to be explained for designers with varying levels of training, such that they can appropriately calibrate their trust in the system~\cite{lee2004trust, krupp2023unreflected}, understand it, and apply it effectively~\cite{haque2023explainable}. 
Second, our findings demonstrate that LLM-based idea evaluation could potentially filter out low-rated ideas in early stages of the process. This is promising, since teams of future or novice designers  could receive early feedback, which provides direction and allows them to focus their time on developing the more promising ideas. Finally, as van Dijk warns us non-human agent still embody human biases~\cite{vanDijk2020}, before making an LLM-based idea evaluation engine available to users, it is important to probe for and identify potential biases in its output.

\subsection{Limitations}
A clear limitation of our work is that we only examined the use of LLMs in ideation with novice designers, using a specific ideation process (Brainwriting), using a single problem statement, and within the context of HCI education. 
 The students were also all novice users of GPT-3. Therefore, the study may not generalize to cases where the groups consist of expert LLM users, expert designers, or users that are assisted by highly trained prompt engineers. Furthermore, the study may not generalize to cases where the participants themselves are experts in the innovative domain, to different innovation domains or to other educational disciplines. Another limitation is that our work lacks an exploration of the long-term impact of integrating AI into HCI education, focusing primarily on immediate outcomes. Nevertheless, our study demonstrates the feasibility of enhancing enhancing Brainwriting with LLMs, and open avenues for future work in the intersection of AI, HCI and education, including developing custom interfaces and conducting longitudinal studies.




\section{Conclusion}
We expect that collaboration between humans and LLMs is one of several radical changes in the way in which humans will utilize machines in the coming years (cf.~\cite{janssen2019history}). In this work we explore one potential scenario of such a collaboration, when an LLM supports a collaborative ideation process of a team. Our focus is on Brainwriting, and we explore how an LLM can enhance the ideas generated by the team using Brainwriting within an educational context, as well as how it can help broaden the number of topics that are explored by the team. Our results indicate that LLMs can be useful in both aspects. Furthermore, we found that LLM-based idea evaluations hold promise in identifying both good ideas and poor ideas, which in the future could be useful feedback to teams as they work through the Brainwriting process, with the caveat that the system must be carefully designed such that its feedback is explainable and avoids propagating biases derived from human-generated data. 
 
\begin{acks} 
This work was in part supported by NSF grant CMMI-1840085. The authors are grateful to Marios Constantinides and Duncan Brumby for generously contributing their time in our early conversations exploring the use of LLMs in group ideation. We also thank Marysabel Morales and Josephine Ramirez for assisting with early explorations of the data.
\end{acks} 
\bibliographystyle{ACM-Reference-Format}
\bibliography{main}
\newpage
\end{document}